\documentclass[a4paper]{article}

\usepackage[utf8]{inputenc}
\usepackage{authblk}
\usepackage{amsmath}
\usepackage{amsthm}
\usepackage{bbold}
\usepackage{enumitem}
\usepackage[autosize]{dot2texi}
\usepackage{tikz}
\usetikzlibrary{shapes,arrows}
\usepackage{subcaption}
\usepackage{xcolor}
\usepackage[toc]{appendix}
\usepackage{hyperref}
\usepackage{graphicx}
\usepackage{floatpag}

\title{Evaluation of Cluster Id Assignment Schemes with ABCDE}
\author[1]{Stephan van Staden}
\affil[1]{Google Switzerland GmbH}
\date{September 2024}

\newcommand{\Base}{\mathit{Base}}
\newcommand{\Exp}{\mathit{Exp}}
\newcommand{\Ideal}{\mathit{Ideal}}
\newcommand{\weight}{\mathit{weight}}
\newcommand{\JaccardDistance}{\mathit{JaccardDistance}}

\newcommand{\SplitRate}{\mathit{SplitRate}}
\newcommand{\MergeRate}{\mathit{MergeRate}}

\newcommand{\GoodSplitRate}{\mathit{GoodSplitRate}}
\newcommand{\BadSplitRate}{\mathit{BadSplitRate}}
\newcommand{\GoodMergeRate}{\mathit{GoodMergeRate}}
\newcommand{\BadMergeRate}{\mathit{BadMergeRate}}

\newcommand{\DeltaPrecision}{\Delta\mathit{Precision}}
\newcommand{\DeltaRecall}{\Delta\mathit{Recall}}

\newcommand{\match}{\equiv}
\newcommand{\distinct}{\not\equiv}
\newcommand{\IQ}{\mathit{IQ}}

\newcommand{\Ids}{\mathit{Ids}}
\newcommand{\baseline}{\mathrm{base}}
\newcommand{\experiment}{\mathrm{exp}}
\newcommand{\hist}{\mathrm{hist}}
\newcommand{\AllIds}{\mathit{AllIds}}
\newcommand{\NonHistIds}{\mathit{NonHistIds}}
\newcommand{\Cluster}{\mathit{Cluster}}
\newcommand{\HistMembersOrId}{\mathit{HistMembersOrId}}
\newcommand{\Weight}{\mathit{Weight}}

\begin{document}

\maketitle

\begin{abstract}
A cluster id assignment scheme labels each cluster of a clustering with a distinct id. The goal of id assignment is semantic id stability, which means that, whenever possible, a cluster for the same underlying concept as that of a historical cluster should ideally receive the same id as the historical cluster. Semantic id stability allows the users of a clustering to refer to a concept's cluster with an id that is stable across clusterings/time. This paper treats the problem of evaluating the relative merits of id assignment schemes. In particular, it considers a historical clustering with id assignments, and a new clustering with ids assigned by a baseline and an experiment. It produces metrics that characterize both the magnitude and the quality of the id assignment diffs between the baseline and the experiment. That happens by transforming the problem of cluster id assignment into a problem of cluster membership, and evaluating it with ABCDE. ABCDE is a sophisticated and scalable technique for evaluating differences in cluster membership in real-world applications, where billions of items are grouped into millions of clusters, and some items are more important than others. The paper also describes several generalizations to the basic evaluation setup for id assignment schemes. For example, it is fairly straightforward to evaluate changes that simultaneously mutate cluster memberships and cluster ids. The ideas are generously illustrated with examples.
\end{abstract}


\section{Introduction}

Clustering is the partitioning of a set of items into separate groups, called clusters. The items in each cluster should typically be similar, while the items from different clusters should be different.

A cluster id assignment scheme is a function that takes as input a clustering, plus additional information, and it produces as output a mapping that associates each cluster with an id that is not shared with any other cluster. The ``additional information'' that id assignment schemes use can differ widely between different schemes.

Each cluster has a semantic identity, which is the main concept or notion of item similarity that it captures. If that identity is also present in a previous clustering, and was associated with an id $id$, then preferably the id assignment scheme should also associate the current cluster with $id$. This property is called ``semantic id stability''. Semantic id stability is useful in practice because it allows users to refer to a specific concept across clusterings with a fixed id. For example, if we cluster geographical information, and the cluster whose items are about Uganda is labeled with the id $\mathit{uganda}$, then users can ask for the latest cluster with the id $\mathit{uganda}$ to get the latest information about Uganda.

In practice, some id assignment schemes attempt to approximate semantic id stability with notions of syntactic id stability, where for example the members of a new cluster can ``vote'' for their previous ids and the id of a cluster is the majority vote (some conflict resolution strategies are employed when multiple clusters want the same id). Other schemes attempt to be more ``semantic'' by employing mechanisms like embedding the clusters into a high-dimensional vector space, and looking which new cluster is closest to the embedding vector of a previous cluster\footnote{One can impose a maximum distance threshold between a previous cluster $P$ and a new cluster $N$, and a minimum distance threshold between $P$ and any other new cluster $N^\prime \neq N$, for $N$ to receive the same id as $P$. All new clusters that did not receive an id in this way will get fresh ids.}.
Id assignment is challenging because the previous clustering can be significantly different from the current clustering. Moreover, the data of individual items can also differ significantly from before and can affect their semantic identities.

The existence of multiple id assignment schemes raises the problem of evaluation. Can we characterize schemes with metrics that highlight their differences, and ultimately help us to decide which scheme to use in practice in a given application?

In the basic evaluation setup, we have:
\begin{itemize}
    \item A historical clustering with ids, which informs us about the semantic identity of ids that were used in the past.
    \item A new clustering.
    \item Ids assigned to the new clustering by two id assignment schemes -- the baseline and the experiment.
\end{itemize}
The goal of the evaluation is to highlight:
\begin{itemize}
    \item The magnitude of the id assignment diffs between the baseline and the experiment. \\
    I.e. is there a big or a small difference between the id assignments of the baseline and the experiment?
    \item The quality of the id assignment diffs between the baseline and the experiment. \\
    I.e. are the diffs in id assignment simply churn (quality-neutral), or are they regressions or improvements with respect to semantic identity?
\end{itemize}
Moreover, the evaluation should take into account that some semantic identities are more important than others, and hence that not all wins/losses are equal.

That seems like a daunting problem. The main contribution of this paper is to demonstrate that it can be tackled with ABCDE~\cite{vanstadengrubb2024abcde,vanstaden2024clusteringqualitymetricsabcde}, a sophisticated evaluation technique for cluster membership changes. On the surface, ABCDE might seem unfit to evaluate cluster id assignments, because it evaluates changes in cluster membership, while a cluster id assignment scheme keeps the members where they are and simply labels the clusters with ids, which is clearly a different problem. However, this discrepancy is not as fundamental as it seems at first glance.

There exists a clear dependency between cluster membership decisions and cluster id assignment decisions.
Without good cluster memberships, i.e. clusters with relatively homogeneous and complete semantic identities, even the best id assignment scheme cannot really do a great job. However, a bad id assignment scheme can completely destroy semantic id stability even when the clusters themselves are perfect. For example, the trivial id assignment scheme that always assigns a fresh id to each cluster will maintain no semantic id stability whatsoever.

This paper also shows that it is not strictly necessary to evaluate changes in cluster membership and changes in cluster id assignment separately. In fact, it is possible to evaluate algorithms that simultaneously change cluster memberships and cluster ids. So algorithms that take as input a clustering labeled with ids, and outputs a different clustering with ids, where both cluster members and cluster ids can change, can be evaluated effectively.

This paper also presents several other generalizations of the basic evaluation setup described above, for example the ability to consider multiple historical clusterings with ids, or the case when the baseline and the experiment assign ids to separate new clusterings instead of a single new clustering.

\section{Basic evaluation}

Suppose we have a baseline id assignment scheme, and an experiment id assignment scheme.
We want to evaluate them against each other for the same clustering. For evaluation purposes we will use a historical clustering where each cluster is labeled with an id.

So we have:
\begin{itemize}
    \item A current clustering $C$, with ids assigned by the baseline, and ids assigned by the experiment.
    \item A historical clustering $H$, with ids assigned by some means in the past.
\end{itemize}

We will treat the items of the current clustering $C$ and the historical clustering $H$ as completely separate, and let $i$ range over current and historical items.

Each clustering associates each of its items with a positive real weight, which encodes the relative importance of the item\footnote{See Section 3.1 of~\cite{vanstadengrubb2024abcde} for more information about item weights.}. So $C$ and $H$ have associated weight functions with disjoint domains, which we can combine in a single function $\weight(i)$ which records weights for all current and historical items.


In the following notation, let $X \in \{\baseline, \experiment, \hist\}$:
\begin{itemize}
\item $\Ids_X$ denotes the set of ids assigned by $X$ to its clustering. \\
      So: \\
      $\Ids_\baseline$ denotes the set of ids assigned by the baseline to clustering $C$. \\
      $\Ids_\experiment$ denotes the set of ids assigned by the experiment to clustering $C$. \\
      $\Ids_\hist$ denotes the set of ids assigned in the past to clustering $H$.
\item $\Cluster_X(id)$ denotes the cluster (i.e. set of items) that $X$ associated with $id$. If $id \notin \Ids_X$, then $\Cluster_X(id) = \emptyset$. \\
Note that $\Cluster_\hist(id)$ is always disjoint from $\Cluster_\baseline(id)$ and $\Cluster_\experiment(id)$ because of the separation of current and historical items, while $\Cluster_\baseline(id)$ and $\Cluster_\experiment(id)$ are not necessarily disjoint.
\item $\AllIds = \Ids_\baseline \cup \Ids_\experiment \cup \Ids_\hist$
\item $\NonHistIds = \AllIds \setminus \Ids_\hist$
\item For $id \in \AllIds$, define
$$\HistMembersOrId(id) = \Cluster_\hist(id) \text{ if } \Cluster_\hist(id) \neq \emptyset \text{ else } \{id\}$$
\end{itemize}

The heart of the basic evaluation consists of crafting inputs for ABCDE, which happens as follows:
\begin{itemize}
\item $\Base$ is a clustering, where for each $id \in \AllIds$ we create a cluster whose members are:
      $$\HistMembersOrId(id) \cup \Cluster_\baseline(id)$$
\item $\Exp$ is a clustering,  where for each $id \in \AllIds$ we create a cluster whose members are:
      $$\HistMembersOrId(id) \cup \Cluster_\experiment(id)$$
\item Provide a weight mapping $\Weight$ that:
\begin{itemize}
    \item Associates each current or historical item $i$ with $\weight(i)$ as described above.
    \item Associates each $id \in \NonHistIds$ with a constant value $k$ that is appropriate for the application at hand. \\ For example, use a value $k$ that would be a reasonable minimum weight for a cluster with no importance/popularity information. Then for every $id \in \Ids_\hist$ we have $\weight(\Cluster_\hist(id)) \geq k$, irrespective of whether the cluster has importance/popularity information\footnote{The $\weight$ of a set of items is simply the sum of the weights of its members.}. As we will see later in the discussion on quality metrics, it ensures that assigning a totally wrong historical id to a cluster is never better than using a fresh id.
\end{itemize}
\end{itemize}

The basic evaluation proceeds by applying ABCDE as usual with inputs $\Base$, $\Exp$ and $\Weight$. Notice that $\Base$ and $\Exp$ are now clusterings of items that are $\NonHistIds$ or current or historical items, so their notion of item is an expanded one. As usual, ABCDE will produce impact and quality metrics.

\subsection{Impact metrics}\label{impact_metrics}
The overall $\JaccardDistance$ will measure the magnitude of the diff in cluster id assignments between the baseline and experiment. We will also have overall $\SplitRate$ and $\MergeRate$ metrics, which characterize how the experiment respectively discarded and adopted historical ids compared to the baseline.

\subsubsection{Example}
Assume that the items and their data stayed the same. Assume $\Ids_\hist = \Ids_\baseline$, and that, modulo the distinction of historical and current items, $H = C$ and $\Cluster_\hist(id) = \Cluster_\baseline(id)$. Assume the experiment assigns brand new cluster ids to all clusters. The ABCDE metrics will show that the $\JaccardDistance$ between the $\Base$ and $\Exp$ clusterings is large. Moreover, the $\SplitRate$ will be much larger than the $\MergeRate$, which shows that many historical items (members of clusters with historical ids, which are heavier than the new ids) were split off from clusters in $\Exp$, i.e. historical ids were dissociated from their clusters in $\Exp$. The $\MergeRate$ will be smaller but still positive, because new ids (which are lighter than the old ids' items) were merged into the clusters in $\Exp$, i.e. newly associated with the clusters.

Strawman experiments like this can provide us with useful calibrations of the metrics, i.e. what we can expect the $\JaccardDistance$/$\SplitRate$/$\MergeRate$ to be in handcrafted situations for the application at hand. Then we can get a good feel for the expected ranges of the metrics and can interpret their magnitudes better for actual experiments.

A toy instance of this example is shown in Figure~\ref{figure_exp_uses_only_fresh_ids}. The historical clustering with id assignments is shown in (a). Historical items are named $h_1$, $h_2$, etc. and have a square shape. Items are grouped together in clusters, with the cluster id shown inside the cluster at the top. Parts (b) and (c) respectively show the baseline and experiment clusterings with id assignments. Their clusters contain current items, which are named $i_1$, $i_2$, etc. and which have a round shape. Parts (d) and (e) show the $\Base$ and the $\Exp$ clusterings that are the inputs to ABCDE. Cluster members that are ids have a diamond shape. Part (f) shows the $\Ideal$ clustering, which will play a role in the next section when we consider the quality; each cluster in $\Ideal$ contains members with a shared unique color, so the color of an item indicates its ideal equivalence class. The $\Weight$ mapping that is used for the metrics is shown in (g), and (h) shows the resulting Impact metrics.

\begin{figure*}[ht!]
    \centering
    \begin{subfigure}[t]{\textwidth}
        \centering
        \begin{dot2tex}[dot,mathmode]
          graph {
            subgraph cluster_id_1 {
              label="id_1"
              style=rounded
              node [color=yellow, style=filled, shape=square]
              h_2 [label=h_2]
              node [color=yellow, style=filled, shape=square]
              h_1 [label=h_1]
            }
            subgraph cluster_id_2 {
              label="id_2"
              style=rounded
              node [color=cyan, style=filled, shape=square]
              h_3 [label=h_3]
            }
          }
        \end{dot2tex}
        \caption{The $\hist$ clustering with cluster ids.}
    \end{subfigure}
    ~
    \par\bigskip
    \begin{subfigure}[b]{0.5\textwidth}
        \centering
        \begin{dot2tex}[dot,mathmode]
          graph {
            subgraph cluster_id_1 {
              label="id_1"
              style=rounded
              node [color=yellow, style=filled, shape=circle]
              i_2 [label=i_2]
              node [color=yellow, style=filled, shape=circle]
              i_1 [label=i_1]
            }
            subgraph cluster_id_2 {
              label="id_2"
              style=rounded
              node [color=cyan, style=filled, shape=circle]
              i_3 [label=i_3]
            }
          }
        \end{dot2tex}
        \caption{The $\baseline$ clustering with cluster ids.}
    \end{subfigure}%
    ~
    \begin{subfigure}[b]{0.5\textwidth}
        \centering
        \begin{dot2tex}[dot,mathmode]
          graph {
            subgraph cluster_id_3 {
              label="id_3"
              style=rounded
              node [color=yellow, style=filled, shape=circle]
              i_2 [label=i_2]
              node [color=yellow, style=filled, shape=circle]
              i_1 [label=i_1]
            }
            subgraph cluster_id_4 {
              label="id_4"
              style=rounded
              node [color=cyan, style=filled, shape=circle]
              i_3 [label=i_3]
            }
          }
        \end{dot2tex}
        \caption{The $\experiment$ clustering with cluster ids.}
    \end{subfigure}
    ~
    \par\bigskip
    \begin{subfigure}[t]{\textwidth}
        \centering
        \begin{dot2tex}[dot,mathmode]
          graph {
            subgraph cluster_0 {
              style=rounded
              node [color=yellow, style=filled, shape=circle]
              i_2 [label=i_2]
              node [color=yellow, style=filled, shape=circle]
              i_1 [label=i_1]
              node [color=yellow, style=filled, shape=square]
              h_2 [label=h_2]
              node [color=yellow, style=filled, shape=square]
              h_1 [label=h_1]
            }
            subgraph cluster_1 {
              style=rounded
              node [color=cyan, style=filled, shape=circle]
              i_3 [label=i_3]
              node [color=cyan, style=filled, shape=square]
              h_3 [label=h_3]
            }
            subgraph cluster_2 {
              style=rounded
              node [color=magenta, style=filled, shape=diamond]
              id_3 [label=id_3]
            }
            subgraph cluster_3 {
              style=rounded
              node [color=red, style=filled, shape=diamond]
              id_4 [label=id_4]
            }
          }
        \end{dot2tex}
        \caption{The $\Base$ clustering.}
    \end{subfigure}
    ~
    \par\bigskip
    \begin{subfigure}[t]{\textwidth}
        \centering
        \begin{dot2tex}[dot,mathmode]
          graph {
            subgraph cluster_0 {
              style=rounded
              node [color=yellow, style=filled, shape=square]
              h_2 [label=h_2]
              node [color=yellow, style=filled, shape=square]
              h_1 [label=h_1]
            }
            subgraph cluster_1 {
              style=rounded
              node [color=cyan, style=filled, shape=square]
              h_3 [label=h_3]
            }
            subgraph cluster_2 {
              style=rounded
              node [color=magenta, style=filled, shape=diamond]
              id_3 [label=id_3]
              node [color=yellow, style=filled, shape=circle]
              i_2 [label=i_2]
              node [color=yellow, style=filled, shape=circle]
              i_1 [label=i_1]
            }
            subgraph cluster_3 {
              style=rounded
              node [color=red, style=filled, shape=diamond]
              id_4 [label=id_4]
              node [color=cyan, style=filled, shape=circle]
              i_3 [label=i_3]
            }
          }
        \end{dot2tex}
        \caption{The $\Exp$ clustering.}
    \end{subfigure}
    ~
    \par\bigskip
    \begin{subfigure}[t]{\textwidth}
        \centering
        \begin{dot2tex}[dot,mathmode]
          graph {
            subgraph cluster_0 {
              style=rounded
              node [color=yellow, style=filled, shape=circle]
              i_2 [label=i_2]
              node [color=yellow, style=filled, shape=circle]
              i_1 [label=i_1]
              node [color=yellow, style=filled, shape=square]
              h_2 [label=h_2]
              node [color=yellow, style=filled, shape=square]
              h_1 [label=h_1]
            }
            subgraph cluster_1 {
              style=rounded
              node [color=cyan, style=filled, shape=circle]
              i_3 [label=i_3]
              node [color=cyan, style=filled, shape=square]
              h_3 [label=h_3]
            }
            subgraph cluster_2 {
              style=rounded
              node [color=magenta, style=filled, shape=diamond]
              id_3 [label=id_3]
            }
            subgraph cluster_3 {
              style=rounded
              node [color=red, style=filled, shape=diamond]
              id_4 [label=id_4]
            }
          }
        \end{dot2tex}
        \caption{The $\Ideal$ clustering.}
    \end{subfigure}
    ~
    \par\bigskip
    \makebox[\linewidth][c]{%
    \begin{subfigure}[b]{0.3666666666666667\textwidth}
        \centering
        \begin{tabular}{|l|l|}
          \hline
          $h_1$ & 1.0 \\
          \hline
          $h_2$ & 1.0 \\
          \hline
          $h_3$ & 1.0 \\
          \hline
          $i_1$ & 1.0 \\
          \hline
          $i_2$ & 1.0 \\
          \hline
          $i_3$ & 1.0 \\
          \hline
          $id_3$ & 0.001 \\
          \hline
          $id_4$ & 0.001 \\
          \hline
        \end{tabular}
        \caption{The $\Weight$ mapping.}
    \end{subfigure}%
    ~
    \begin{subfigure}[b]{0.3666666666666667\textwidth}
        \centering
        \begin{tabular}{|l|r|}
          \hline
          $\JaccardDistance$ & 50.02\% \\
          \hline
          $\SplitRate$ & 49.98\% \\
          \hline
          $\MergeRate$ & 0.07\% \\
          \hline
        \end{tabular}
        \caption{Impact metrics.}
    \end{subfigure}%
    ~
    \begin{subfigure}[b]{0.3666666666666667\textwidth}
        \centering
        \begin{tabular}{|l|r|}
          \hline
          $\GoodSplitRate$ & 0.00\% \\
          \hline
          $\BadSplitRate$ & 49.98\% \\
          \hline
          $\GoodMergeRate$ & 0.00\% \\
          \hline
          $\BadMergeRate$ & 0.07\% \\
          \hline
          $\DeltaPrecision$ & -0.07\% \\
          \hline
          $\DeltaRecall$ & -49.98\% \\
          \hline
          $\IQ$ & -100.00\% \\
          \hline
        \end{tabular}
        \caption{Quality metrics.}
    \end{subfigure}
    }
    \caption{The experiment assigns fresh ids to all clusters.}\label{figure_exp_uses_only_fresh_ids}
\end{figure*}

\subsection{Quality metrics}

\subsubsection{Human judgements for quality metrics}

When it comes to the quality of the diff in cluster id assignments between the baseline and the experiment, the usual ABCDE sampling will yield pairs for human judgement. There are two kinds of pairs:
\begin{itemize}
\item Pairs that consist of two items. \\ Such pairs can be judged as usual: for a pair $(i_1, i_2)$, humans can say whether $i_1 \match i_2$ or $i_1 \distinct i_2$, i.e. whether they are similar or distinct\footnote{Pairs for which humans are uncertain, or cannot reach a verdict, are discarded by ABCDE~when computing quality metrics; the remaining pairs are reweighted to avoid bias as described in bullet 4 of Section 5.8 in~\cite{vanstadengrubb2024abcde}.}. \\
Note that these pairs can involve historical and current items. Nonetheless, they are still questions about whether items have the same underlying identity or not, so there are no conceptual changes to the judgement task.
\item Pairs that consist of an item and an id, where the id is a member of $\NonHistIds$. Such pairs have the form $(i, id)$ or $(id, i)$. \\
There is no need to send these to humans, since they should always receive the distinct verdict $i \distinct id$. \\
Rationale:
Intuitively, the question is whether the item should be associated with the id on the basis of the id's historical semantics. Since the id has no historical semantics, there is no evidence to justify a $i \match id$ verdict. This handling will also cause the quality metrics to award a small penalty for assigning a fresh id to a cluster, but a proper choice of $k$ will ensure that the penalty is not larger than the penalty for using a totally wrong historical id.
\end{itemize}

\subsubsection{Interpreting the ABCDE quality metrics}

In Section~\ref{impact_metrics} on Impact metrics, we saw that the $\SplitRate$ and $\MergeRate$ metrics characterize how the experiment respectively discarded and adopted historical ids compared to the baseline. The quality metrics will decompose each of them further to express the good and the bad components. In particular, we have:
\begin{align*}
\SplitRate &= \GoodSplitRate + \BadSplitRate \\
\MergeRate &= \GoodMergeRate + \BadMergeRate
\end{align*}
The $\GoodSplitRate$ measures the historical ids that the experiment correctly dissociated (compared to the baseline), while the $\BadSplitRate$ measures the historical ids that the experiment incorrectly dissociated. Similarly, the $\GoodMergeRate$ denotes the degree to which the experiment correctly adopted historical ids compared to the baseline, and the $\BadMergeRate$ the degree to which the experiment incorrectly adopted historical ids.

The $\DeltaPrecision$ metric characterizes the degree to which the experiment improved the homogeneity of the semantic identities when assigning historical ids to current clusters. For example, switching around the historical labels of the clusters for Uganda and Zambia in the experiment will lead to a loss in precision, and hence to a negative value for $\DeltaPrecision$.

The $\DeltaRecall$ metric characterizes the degree to which the experiment improved the completeness of the semantic identities when assigning historical ids to current clusters. For example, dropping the historical label of the cluster for Uganda in the experiment, and assigning it a fresh id instead, will lead to a drop in recall and hence to a negative value for $\DeltaRecall$.

Finally, the $\IQ$ metric characterizes the relationship between the Impact and the Quality of the change. It expresses the degree to which the difference in the id assignments between the baseline and the experiment translates into an improvement in the quality of the id assignments. Formally, $\IQ = X\%$ means that for every 100 units of $\JaccardDistance$ between $\Base$ and $\Exp$, the $\Exp$ clustering is $X$ units of $\JaccardDistance$ closer to the $\Ideal$ clustering. The best possible value for $\IQ$ is $100\%$, while the worst is $-100\%$, and a quality-neutral change will have $\IQ = 0\%$.

\subsubsection{Examples}

In a real-world application, with billions of items and millions of clusters, the $\Ideal$ clustering is not fully known (if it was fully known, then there would be no point to experiment with id assignment schemes). ABCDE will sample item pairs for human judgement, which is an on-demand expansion of the $\Ideal$ clustering to a practical but limited extent, and then ABCDE uses the outcomes of the human judgements to estimate the quality metrics. 

In contrast to that, each toy example of this paper comes with a fully materialized $\Ideal$ clustering, and the quality metrics are computed exactly. So there are no stochastic effects in the quality metrics of the examples, which is great for illustration purposes. The $\Ideal$ clustering partitions all the elements according to their ideal equivalence classes. An element can be a historical item, a current item, or a member of $\NonHistIds$. To make it easier to understand, the color of each element in an example denotes its ideal equivalence class, and the shape of an item denotes its kind (historical items are squares, current items are circles, ids are diamonds). Here are the toy examples:

\begin{enumerate}
\item Figure~\ref{figure_exp_uses_only_fresh_ids} shows that dropping historical ids, and using fresh ids instead, will lead to a large $\BadSplitRate$ and a large loss of recall (and hence a large negative value for $\DeltaRecall$). It will also lead to a small negative value for $\DeltaPrecision$, which is the penalty for introducing fresh ids. The fact that $\IQ$ has the minimum possible value says that the experiment is messing up the quality of id assignment, which was perfect in the baseline.
\item Figure~\ref{figure_exp_swaps_ids} is similar, but the experiments swaps the historical ids instead of dropping them. The dissociation of the historical ids with their clusters causes the large $\BadSplitRate$ and the large negative $\DeltaRecall$. The association of the historical ids with the wrong clusters causes the large $\BadMergeRate$ and the large negative $\DeltaPrecision$. Again, the $\IQ$ tells us that the experiment is messing up the quality of id assignment.
\item Figure~\ref{figure_assigning_ids_when_cluster_split} shows what happens if a historical cluster that was perfectly homogeneous is split up in the current clustering. If the identities of the items remained the same, then it is best to assign the historical id to the current cluster with the largest weight. That is reflected in the quality metrics by the positive values for $\DeltaRecall$ and $\IQ$.
\item Figure~\ref{figure_historical_id_with_ambiguous_meaning} shows a conflated historical cluster. The historical id therefore has an ambiguous semantics -- two thirds of it stand for blue and one third for yellow. When the blue items are absent from the current clustering, then there are two obvious things to do with the current cluster with the yellow item: either assign it a fresh id (as the baseline does), or assign it the historical id (as the experiment does). The quality metrics punish the experiment for assigning a mostly-blue id to a yellow item: $\DeltaPrecision$ is clearly negative. But they also reward the experiment for remembering that the historical id was also associated with a non-trivial amount of yellow: the $\DeltaRecall$ metric is clearly positive. The $\IQ$ metric says that the experiment is quality-negative, in the sense that it is further away from the ideal situation compared to the baseline. So using a fresh id for the current cluster is more appropriate.
\item Figure~\ref{figure_historical_id_for_conflated_cluster} shows the same conflated historical cluster as in the previous example, but there is also a homogeneous magenta historical cluster. The current clustering has a single cluster that contains the yellow and magenta items; the blue items are absent. The example shows that, if we are forced to pick a historical id for the new cluster, then it is better to pick the magenta id, since the conflated id is burdened by its majority-blue semantics, which would cause a loss of precision.
\item Figure~\ref{figure_fresh_id_for_conflated_cluster} has a similar historical and baseline situation, but the experiment uses a fresh id for the conflated cluster. The quality metrics show an improvement in precision, because the magenta id is not associated with yellow items in the experiment. However, there is a clear loss of recall, because the magenta item is not associated with the magenta id in the experiment. The negative value of $\IQ$ means that the experiment, which basically discards all historical ids, is quality-negative because it is further away from $\Ideal$ when compared to the baseline.
\item Figure~\ref{figure_fresh_id_is_better_than_important_historical_id} shows that using a fresh id for a conflated cluster can be better than using an important historical id. Notice that $\weight(h_1) = 10.0$, which makes $id_1$ relatively important. The example is somewhat exaggerated: a value of $\weight(h_1) = 1.1$ already yields a positive $\IQ$. 
\item Figure~\ref{figure_better_to_use_fresh_id_than_wrong_id} shows that, instead of using a totally wrong historical id, it is better to use a fresh id.
\item Figure~\ref{figure_base_and_exp_use_only_fresh_ids} shows a baseline and an experiment that operate similarly: they discard all historical ids and assign fresh ids to all clusters. The main quality metrics, namely $\DeltaPrecision$, $\DeltaRecall$ and $\IQ$, are all neutral. The churn in ids is reflected in the impact metrics and also in the fact that $\GoodSplitRate = \BadMergeRate$. Notice that, if the baseline and the experiment used exactly the same fresh ids for the same clusters, then all the metrics would be neutral.
\end{enumerate}

\begin{figure*}[ht!]
    \centering
    \begin{subfigure}[t]{\textwidth}
        \centering
        \begin{dot2tex}[dot,mathmode]
          graph {
            subgraph cluster_id_1 {
              label="id_1"
              style=rounded
              node [color=yellow, style=filled, shape=square]
              h_2 [label=h_2]
              node [color=yellow, style=filled, shape=square]
              h_1 [label=h_1]
            }
            subgraph cluster_id_2 {
              label="id_2"
              style=rounded
              node [color=cyan, style=filled, shape=square]
              h_3 [label=h_3]
            }
          }
        \end{dot2tex}
        \caption{The $\hist$ clustering with cluster ids.}
    \end{subfigure}
    ~
    \par\bigskip
    \begin{subfigure}[b]{0.5\textwidth}
        \centering
        \begin{dot2tex}[dot,mathmode]
          graph {
            subgraph cluster_id_1 {
              label="id_1"
              style=rounded
              node [color=yellow, style=filled, shape=circle]
              i_2 [label=i_2]
              node [color=yellow, style=filled, shape=circle]
              i_1 [label=i_1]
            }
            subgraph cluster_id_2 {
              label="id_2"
              style=rounded
              node [color=cyan, style=filled, shape=circle]
              i_3 [label=i_3]
            }
          }
        \end{dot2tex}
        \caption{The $\baseline$ clustering with cluster ids.}
    \end{subfigure}%
    ~
    \begin{subfigure}[b]{0.5\textwidth}
        \centering
        \begin{dot2tex}[dot,mathmode]
          graph {
            subgraph cluster_id_2 {
              label="id_2"
              style=rounded
              node [color=yellow, style=filled, shape=circle]
              i_2 [label=i_2]
              node [color=yellow, style=filled, shape=circle]
              i_1 [label=i_1]
            }
            subgraph cluster_id_1 {
              label="id_1"
              style=rounded
              node [color=cyan, style=filled, shape=circle]
              i_3 [label=i_3]
            }
          }
        \end{dot2tex}
        \caption{The $\experiment$ clustering with cluster ids.}
    \end{subfigure}
    ~
    \par\bigskip
    \begin{subfigure}[t]{\textwidth}
        \centering
        \begin{dot2tex}[dot,mathmode]
          graph {
            subgraph cluster_0 {
              style=rounded
              node [color=yellow, style=filled, shape=circle]
              i_2 [label=i_2]
              node [color=yellow, style=filled, shape=circle]
              i_1 [label=i_1]
              node [color=yellow, style=filled, shape=square]
              h_2 [label=h_2]
              node [color=yellow, style=filled, shape=square]
              h_1 [label=h_1]
            }
            subgraph cluster_1 {
              style=rounded
              node [color=cyan, style=filled, shape=circle]
              i_3 [label=i_3]
              node [color=cyan, style=filled, shape=square]
              h_3 [label=h_3]
            }
          }
        \end{dot2tex}
        \caption{The $\Base$ clustering.}
    \end{subfigure}
    ~
    \par\bigskip
    \begin{subfigure}[t]{\textwidth}
        \centering
        \begin{dot2tex}[dot,mathmode]
          graph {
            subgraph cluster_0 {
              style=rounded
              node [color=cyan, style=filled, shape=circle]
              i_3 [label=i_3]
              node [color=yellow, style=filled, shape=square]
              h_2 [label=h_2]
              node [color=yellow, style=filled, shape=square]
              h_1 [label=h_1]
            }
            subgraph cluster_1 {
              style=rounded
              node [color=yellow, style=filled, shape=circle]
              i_2 [label=i_2]
              node [color=yellow, style=filled, shape=circle]
              i_1 [label=i_1]
              node [color=cyan, style=filled, shape=square]
              h_3 [label=h_3]
            }
          }
        \end{dot2tex}
        \caption{The $\Exp$ clustering.}
    \end{subfigure}
    ~
    \par\bigskip
    \begin{subfigure}[t]{\textwidth}
        \centering
        \begin{dot2tex}[dot,mathmode]
          graph {
            subgraph cluster_0 {
              style=rounded
              node [color=yellow, style=filled, shape=circle]
              i_2 [label=i_2]
              node [color=yellow, style=filled, shape=circle]
              i_1 [label=i_1]
              node [color=yellow, style=filled, shape=square]
              h_2 [label=h_2]
              node [color=yellow, style=filled, shape=square]
              h_1 [label=h_1]
            }
            subgraph cluster_1 {
              style=rounded
              node [color=cyan, style=filled, shape=circle]
              i_3 [label=i_3]
              node [color=cyan, style=filled, shape=square]
              h_3 [label=h_3]
            }
          }
        \end{dot2tex}
        \caption{The $\Ideal$ clustering.}
    \end{subfigure}
    ~
    \par\bigskip
    \makebox[\linewidth][c]{%
    \begin{subfigure}[b]{0.3666666666666667\textwidth}
        \centering
        \begin{tabular}{|l|l|}
          \hline
          $h_1$ & 1.0 \\
          \hline
          $h_2$ & 1.0 \\
          \hline
          $h_3$ & 1.0 \\
          \hline
          $i_1$ & 1.0 \\
          \hline
          $i_2$ & 1.0 \\
          \hline
          $i_3$ & 1.0 \\
          \hline
        \end{tabular}
        \caption{The $\Weight$ mapping.}
    \end{subfigure}%
    ~
    \begin{subfigure}[b]{0.3666666666666667\textwidth}
        \centering
        \begin{tabular}{|l|r|}
          \hline
          $\JaccardDistance$ & 65.00\% \\
          \hline
          $\SplitRate$ & 50.00\% \\
          \hline
          $\MergeRate$ & 44.44\% \\
          \hline
        \end{tabular}
        \caption{Impact metrics.}
    \end{subfigure}%
    ~
    \begin{subfigure}[b]{0.3666666666666667\textwidth}
        \centering
        \begin{tabular}{|l|r|}
          \hline
          $\GoodSplitRate$ & 0.00\% \\
          \hline
          $\BadSplitRate$ & 50.00\% \\
          \hline
          $\GoodMergeRate$ & 0.00\% \\
          \hline
          $\BadMergeRate$ & 44.44\% \\
          \hline
          $\DeltaPrecision$ & -44.44\% \\
          \hline
          $\DeltaRecall$ & -50.00\% \\
          \hline
          $\IQ$ & -100.00\% \\
          \hline
        \end{tabular}
        \caption{Quality metrics.}
    \end{subfigure}
    }
    \caption{The experiment swaps ids.}\label{figure_exp_swaps_ids}
\end{figure*}

\begin{figure*}[ht!]
    \centering
    \begin{subfigure}[t]{\textwidth}
        \centering
        \begin{dot2tex}[dot,mathmode]
          graph {
            subgraph cluster_id_1 {
              label="id_1"
              style=rounded
              node [color=yellow, style=filled, shape=square]
              h_3 [label=h_3]
              node [color=yellow, style=filled, shape=square]
              h_2 [label=h_2]
              node [color=yellow, style=filled, shape=square]
              h_1 [label=h_1]
            }
          }
        \end{dot2tex}
        \caption{The $\hist$ clustering with cluster ids.}
    \end{subfigure}
    ~
    \par\bigskip
    \begin{subfigure}[b]{0.5\textwidth}
        \centering
        \begin{dot2tex}[dot,mathmode]
          graph {
            subgraph cluster_id_2 {
              label="id_2"
              style=rounded
              node [color=yellow, style=filled, shape=circle]
              i_2 [label=i_2]
              node [color=yellow, style=filled, shape=circle]
              i_1 [label=i_1]
            }
            subgraph cluster_id_1 {
              label="id_1"
              style=rounded
              node [color=yellow, style=filled, shape=circle]
              i_3 [label=i_3]
            }
          }
        \end{dot2tex}
        \caption{The $\baseline$ clustering with cluster ids.}
    \end{subfigure}%
    ~
    \begin{subfigure}[b]{0.5\textwidth}
        \centering
        \begin{dot2tex}[dot,mathmode]
          graph {
            subgraph cluster_id_1 {
              label="id_1"
              style=rounded
              node [color=yellow, style=filled, shape=circle]
              i_2 [label=i_2]
              node [color=yellow, style=filled, shape=circle]
              i_1 [label=i_1]
            }
            subgraph cluster_id_3 {
              label="id_3"
              style=rounded
              node [color=yellow, style=filled, shape=circle]
              i_3 [label=i_3]
            }
          }
        \end{dot2tex}
        \caption{The $\experiment$ clustering with cluster ids.}
    \end{subfigure}
    ~
    \par\bigskip
    \begin{subfigure}[t]{\textwidth}
        \centering
        \begin{dot2tex}[dot,mathmode]
          graph {
            subgraph cluster_0 {
              style=rounded
              node [color=yellow, style=filled, shape=circle]
              i_3 [label=i_3]
              node [color=yellow, style=filled, shape=square]
              h_3 [label=h_3]
              node [color=yellow, style=filled, shape=square]
              h_2 [label=h_2]
              node [color=yellow, style=filled, shape=square]
              h_1 [label=h_1]
            }
            subgraph cluster_1 {
              style=rounded
              node [color=cyan, style=filled, shape=diamond]
              id_2 [label=id_2]
              node [color=yellow, style=filled, shape=circle]
              i_2 [label=i_2]
              node [color=yellow, style=filled, shape=circle]
              i_1 [label=i_1]
            }
            subgraph cluster_2 {
              style=rounded
              node [color=magenta, style=filled, shape=diamond]
              id_3 [label=id_3]
            }
          }
        \end{dot2tex}
        \caption{The $\Base$ clustering.}
    \end{subfigure}
    ~
    \par\bigskip
    \begin{subfigure}[t]{\textwidth}
        \centering
        \begin{dot2tex}[dot,mathmode]
          graph {
            subgraph cluster_0 {
              style=rounded
              node [color=yellow, style=filled, shape=circle]
              i_2 [label=i_2]
              node [color=yellow, style=filled, shape=circle]
              i_1 [label=i_1]
              node [color=yellow, style=filled, shape=square]
              h_3 [label=h_3]
              node [color=yellow, style=filled, shape=square]
              h_2 [label=h_2]
              node [color=yellow, style=filled, shape=square]
              h_1 [label=h_1]
            }
            subgraph cluster_1 {
              style=rounded
              node [color=magenta, style=filled, shape=diamond]
              id_3 [label=id_3]
              node [color=yellow, style=filled, shape=circle]
              i_3 [label=i_3]
            }
            subgraph cluster_2 {
              style=rounded
              node [color=cyan, style=filled, shape=diamond]
              id_2 [label=id_2]
            }
          }
        \end{dot2tex}
        \caption{The $\Exp$ clustering.}
    \end{subfigure}
    ~
    \par\bigskip
    \begin{subfigure}[t]{\textwidth}
        \centering
        \begin{dot2tex}[dot,mathmode]
          graph {
            subgraph cluster_0 {
              style=rounded
              node [color=yellow, style=filled, shape=circle]
              i_3 [label=i_3]
              node [color=yellow, style=filled, shape=circle]
              i_2 [label=i_2]
              node [color=yellow, style=filled, shape=circle]
              i_1 [label=i_1]
              node [color=yellow, style=filled, shape=square]
              h_3 [label=h_3]
              node [color=yellow, style=filled, shape=square]
              h_2 [label=h_2]
              node [color=yellow, style=filled, shape=square]
              h_1 [label=h_1]
            }
            subgraph cluster_1 {
              style=rounded
              node [color=cyan, style=filled, shape=diamond]
              id_2 [label=id_2]
            }
            subgraph cluster_2 {
              style=rounded
              node [color=magenta, style=filled, shape=diamond]
              id_3 [label=id_3]
            }
          }
        \end{dot2tex}
        \caption{The $\Ideal$ clustering.}
    \end{subfigure}
    ~
    \par\bigskip
    \makebox[\linewidth][c]{%
    \begin{subfigure}[b]{0.3666666666666667\textwidth}
        \centering
        \begin{tabular}{|l|l|}
          \hline
          $h_1$ & 1.0 \\
          \hline
          $h_2$ & 1.0 \\
          \hline
          $h_3$ & 1.0 \\
          \hline
          $i_1$ & 1.0 \\
          \hline
          $i_2$ & 1.0 \\
          \hline
          $i_3$ & 1.0 \\
          \hline
          $id_2$ & 0.001 \\
          \hline
          $id_3$ & 0.001 \\
          \hline
        \end{tabular}
        \caption{The $\Weight$ mapping.}
    \end{subfigure}%
    ~
    \begin{subfigure}[b]{0.3666666666666667\textwidth}
        \centering
        \begin{tabular}{|l|r|}
          \hline
          $\JaccardDistance$ & 57.52\% \\
          \hline
          $\SplitRate$ & 25.02\% \\
          \hline
          $\MergeRate$ & 40.02\% \\
          \hline
        \end{tabular}
        \caption{Impact metrics.}
    \end{subfigure}%
    ~
    \begin{subfigure}[b]{0.3666666666666667\textwidth}
        \centering
        \begin{tabular}{|l|r|}
          \hline
          $\GoodSplitRate$ & 0.03\% \\
          \hline
          $\BadSplitRate$ & 24.99\% \\
          \hline
          $\GoodMergeRate$ & 39.99\% \\
          \hline
          $\BadMergeRate$ & 0.03\% \\
          \hline
          $\DeltaPrecision$ & 0.00\% \\
          \hline
          $\DeltaRecall$ & 16.66\% \\
          \hline
          $\IQ$ & 28.97\% \\
          \hline
        \end{tabular}
        \caption{Quality metrics.}
    \end{subfigure}
    }
    \caption{Assigning ids when a cluster splits.}\label{figure_assigning_ids_when_cluster_split}
\end{figure*}

\begin{figure*}[ht!]
    \centering
    \begin{subfigure}[t]{\textwidth}
        \centering
        \begin{dot2tex}[dot,mathmode]
          graph {
            subgraph cluster_id_1 {
              label="id_1"
              style=rounded
              node [color=cyan, style=filled, shape=square]
              h_3 [label=h_3]
              node [color=cyan, style=filled, shape=square]
              h_2 [label=h_2]
              node [color=yellow, style=filled, shape=square]
              h_1 [label=h_1]
            }
          }
        \end{dot2tex}
        \caption{The $\hist$ clustering with cluster ids.}
    \end{subfigure}
    ~
    \par\bigskip
    \begin{subfigure}[b]{0.5\textwidth}
        \centering
        \begin{dot2tex}[dot,mathmode]
          graph {
            subgraph cluster_id_2 {
              label="id_2"
              style=rounded
              node [color=yellow, style=filled, shape=circle]
              i_1 [label=i_1]
            }
          }
        \end{dot2tex}
        \caption{The $\baseline$ clustering with cluster ids.}
    \end{subfigure}%
    ~
    \begin{subfigure}[b]{0.5\textwidth}
        \centering
        \begin{dot2tex}[dot,mathmode]
          graph {
            subgraph cluster_id_1 {
              label="id_1"
              style=rounded
              node [color=yellow, style=filled, shape=circle]
              i_1 [label=i_1]
            }
          }
        \end{dot2tex}
        \caption{The $\experiment$ clustering with cluster ids.}
    \end{subfigure}
    ~
    \par\bigskip
    \begin{subfigure}[t]{\textwidth}
        \centering
        \begin{dot2tex}[dot,mathmode]
          graph {
            subgraph cluster_0 {
              style=rounded
              node [color=cyan, style=filled, shape=square]
              h_3 [label=h_3]
              node [color=cyan, style=filled, shape=square]
              h_2 [label=h_2]
              node [color=yellow, style=filled, shape=square]
              h_1 [label=h_1]
            }
            subgraph cluster_1 {
              style=rounded
              node [color=magenta, style=filled, shape=diamond]
              id_2 [label=id_2]
              node [color=yellow, style=filled, shape=circle]
              i_1 [label=i_1]
            }
          }
        \end{dot2tex}
        \caption{The $\Base$ clustering.}
    \end{subfigure}
    ~
    \par\bigskip
    \begin{subfigure}[t]{\textwidth}
        \centering
        \begin{dot2tex}[dot,mathmode]
          graph {
            subgraph cluster_0 {
              style=rounded
              node [color=yellow, style=filled, shape=circle]
              i_1 [label=i_1]
              node [color=cyan, style=filled, shape=square]
              h_3 [label=h_3]
              node [color=cyan, style=filled, shape=square]
              h_2 [label=h_2]
              node [color=yellow, style=filled, shape=square]
              h_1 [label=h_1]
            }
            subgraph cluster_1 {
              style=rounded
              node [color=magenta, style=filled, shape=diamond]
              id_2 [label=id_2]
            }
          }
        \end{dot2tex}
        \caption{The $\Exp$ clustering.}
    \end{subfigure}
    ~
    \par\bigskip
    \begin{subfigure}[t]{\textwidth}
        \centering
        \begin{dot2tex}[dot,mathmode]
          graph {
            subgraph cluster_0 {
              style=rounded
              node [color=yellow, style=filled, shape=circle]
              i_1 [label=i_1]
              node [color=yellow, style=filled, shape=square]
              h_1 [label=h_1]
            }
            subgraph cluster_1 {
              style=rounded
              node [color=cyan, style=filled, shape=square]
              h_3 [label=h_3]
              node [color=cyan, style=filled, shape=square]
              h_2 [label=h_2]
            }
            subgraph cluster_2 {
              style=rounded
              node [color=magenta, style=filled, shape=diamond]
              id_2 [label=id_2]
            }
          }
        \end{dot2tex}
        \caption{The $\Ideal$ clustering.}
    \end{subfigure}
    ~
    \par\bigskip
    \makebox[\linewidth][c]{%
    \begin{subfigure}[b]{0.3666666666666667\textwidth}
        \centering
        \begin{tabular}{|l|l|}
          \hline
          $h_1$ & 1.0 \\
          \hline
          $h_2$ & 1.0 \\
          \hline
          $h_3$ & 1.0 \\
          \hline
          $i_1$ & 1.0 \\
          \hline
          $id_2$ & 0.001 \\
          \hline
        \end{tabular}
        \caption{The $\Weight$ mapping.}
    \end{subfigure}%
    ~
    \begin{subfigure}[b]{0.3666666666666667\textwidth}
        \centering
        \begin{tabular}{|l|r|}
          \hline
          $\JaccardDistance$ & 37.52\% \\
          \hline
          $\SplitRate$ & 0.05\% \\
          \hline
          $\MergeRate$ & 37.49\% \\
          \hline
        \end{tabular}
        \caption{Impact metrics.}
    \end{subfigure}%
    ~
    \begin{subfigure}[b]{0.3666666666666667\textwidth}
        \centering
        \begin{tabular}{|l|r|}
          \hline
          $\GoodSplitRate$ & 0.05\% \\
          \hline
          $\BadSplitRate$ & 0.00\% \\
          \hline
          $\GoodMergeRate$ & 12.50\% \\
          \hline
          $\BadMergeRate$ & 24.99\% \\
          \hline
          $\DeltaPrecision$ & -16.61\% \\
          \hline
          $\DeltaRecall$ & 24.99\% \\
          \hline
          $\IQ$ & -5.47\% \\
          \hline
        \end{tabular}
        \caption{Quality metrics.}
    \end{subfigure}
    }
    \caption{A historical id with ambiguous meaning.}\label{figure_historical_id_with_ambiguous_meaning}
\end{figure*}

\begin{figure*}[ht!]
    \centering
    \begin{subfigure}[t]{\textwidth}
        \centering
        \begin{dot2tex}[dot,mathmode]
          graph {
            subgraph cluster_id_1 {
              label="id_1"
              style=rounded
              node [color=cyan, style=filled, shape=square]
              h_3 [label=h_3]
              node [color=cyan, style=filled, shape=square]
              h_2 [label=h_2]
              node [color=yellow, style=filled, shape=square]
              h_1 [label=h_1]
            }
            subgraph cluster_id_2 {
              label="id_2"
              style=rounded
              node [color=magenta, style=filled, shape=square]
              h_4 [label=h_4]
            }
          }
        \end{dot2tex}
        \caption{The $\hist$ clustering with cluster ids.}
    \end{subfigure}
    ~
    \par\bigskip
    \begin{subfigure}[b]{0.5\textwidth}
        \centering
        \begin{dot2tex}[dot,mathmode]
          graph {
            subgraph cluster_id_2 {
              label="id_2"
              style=rounded
              node [color=magenta, style=filled, shape=circle]
              i_4 [label=i_4]
              node [color=yellow, style=filled, shape=circle]
              i_1 [label=i_1]
            }
          }
        \end{dot2tex}
        \caption{The $\baseline$ clustering with cluster ids.}
    \end{subfigure}%
    ~
    \begin{subfigure}[b]{0.5\textwidth}
        \centering
        \begin{dot2tex}[dot,mathmode]
          graph {
            subgraph cluster_id_1 {
              label="id_1"
              style=rounded
              node [color=magenta, style=filled, shape=circle]
              i_4 [label=i_4]
              node [color=yellow, style=filled, shape=circle]
              i_1 [label=i_1]
            }
          }
        \end{dot2tex}
        \caption{The $\experiment$ clustering with cluster ids.}
    \end{subfigure}
    ~
    \par\bigskip
    \begin{subfigure}[t]{\textwidth}
        \centering
        \begin{dot2tex}[dot,mathmode]
          graph {
            subgraph cluster_0 {
              style=rounded
              node [color=cyan, style=filled, shape=square]
              h_3 [label=h_3]
              node [color=cyan, style=filled, shape=square]
              h_2 [label=h_2]
              node [color=yellow, style=filled, shape=square]
              h_1 [label=h_1]
            }
            subgraph cluster_1 {
              style=rounded
              node [color=magenta, style=filled, shape=circle]
              i_4 [label=i_4]
              node [color=yellow, style=filled, shape=circle]
              i_1 [label=i_1]
              node [color=magenta, style=filled, shape=square]
              h_4 [label=h_4]
            }
          }
        \end{dot2tex}
        \caption{The $\Base$ clustering.}
    \end{subfigure}
    ~
    \par\bigskip
    \begin{subfigure}[t]{\textwidth}
        \centering
        \begin{dot2tex}[dot,mathmode]
          graph {
            subgraph cluster_0 {
              style=rounded
              node [color=magenta, style=filled, shape=circle]
              i_4 [label=i_4]
              node [color=yellow, style=filled, shape=circle]
              i_1 [label=i_1]
              node [color=cyan, style=filled, shape=square]
              h_3 [label=h_3]
              node [color=cyan, style=filled, shape=square]
              h_2 [label=h_2]
              node [color=yellow, style=filled, shape=square]
              h_1 [label=h_1]
            }
            subgraph cluster_1 {
              style=rounded
              node [color=magenta, style=filled, shape=square]
              h_4 [label=h_4]
            }
          }
        \end{dot2tex}
        \caption{The $\Exp$ clustering.}
    \end{subfigure}
    ~
    \par\bigskip
    \begin{subfigure}[t]{\textwidth}
        \centering
        \begin{dot2tex}[dot,mathmode]
          graph {
            subgraph cluster_0 {
              style=rounded
              node [color=yellow, style=filled, shape=circle]
              i_1 [label=i_1]
              node [color=yellow, style=filled, shape=square]
              h_1 [label=h_1]
            }
            subgraph cluster_1 {
              style=rounded
              node [color=cyan, style=filled, shape=square]
              h_3 [label=h_3]
              node [color=cyan, style=filled, shape=square]
              h_2 [label=h_2]
            }
            subgraph cluster_2 {
              style=rounded
              node [color=magenta, style=filled, shape=circle]
              i_4 [label=i_4]
              node [color=magenta, style=filled, shape=square]
              h_4 [label=h_4]
            }
          }
        \end{dot2tex}
        \caption{The $\Ideal$ clustering.}
    \end{subfigure}
    ~
    \par\bigskip
    \makebox[\linewidth][c]{%
    \begin{subfigure}[b]{0.3666666666666667\textwidth}
        \centering
        \begin{tabular}{|l|l|}
          \hline
          $h_1$ & 1.0 \\
          \hline
          $h_2$ & 1.0 \\
          \hline
          $h_3$ & 1.0 \\
          \hline
          $h_4$ & 1.0 \\
          \hline
          $i_1$ & 1.0 \\
          \hline
          $i_4$ & 1.0 \\
          \hline
        \end{tabular}
        \caption{The $\Weight$ mapping.}
    \end{subfigure}%
    ~
    \begin{subfigure}[b]{0.3666666666666667\textwidth}
        \centering
        \begin{tabular}{|l|r|}
          \hline
          $\JaccardDistance$ & 53.33\% \\
          \hline
          $\SplitRate$ & 22.22\% \\
          \hline
          $\MergeRate$ & 40.00\% \\
          \hline
        \end{tabular}
        \caption{Impact metrics.}
    \end{subfigure}%
    ~
    \begin{subfigure}[b]{0.3666666666666667\textwidth}
        \centering
        \begin{tabular}{|l|r|}
          \hline
          $\GoodSplitRate$ & 11.11\% \\
          \hline
          $\BadSplitRate$ & 11.11\% \\
          \hline
          $\GoodMergeRate$ & 6.67\% \\
          \hline
          $\BadMergeRate$ & 33.33\% \\
          \hline
          $\DeltaPrecision$ & -8.89\% \\
          \hline
          $\DeltaRecall$ & 0.00\% \\
          \hline
          $\IQ$ & -28.13\% \\
          \hline
        \end{tabular}
        \caption{Quality metrics.}
    \end{subfigure}
    }
    \caption{Using a historical id for a conflated cluster.}\label{figure_historical_id_for_conflated_cluster}
\end{figure*}

\begin{figure*}[ht!]
    \centering
    \begin{subfigure}[t]{\textwidth}
        \centering
        \begin{dot2tex}[dot,mathmode]
          graph {
            subgraph cluster_id_1 {
              label="id_1"
              style=rounded
              node [color=cyan, style=filled, shape=square]
              h_3 [label=h_3]
              node [color=cyan, style=filled, shape=square]
              h_2 [label=h_2]
              node [color=yellow, style=filled, shape=square]
              h_1 [label=h_1]
            }
            subgraph cluster_id_2 {
              label="id_2"
              style=rounded
              node [color=magenta, style=filled, shape=square]
              h_4 [label=h_4]
            }
          }
        \end{dot2tex}
        \caption{The $\hist$ clustering with cluster ids.}
    \end{subfigure}
    ~
    \par\bigskip
    \begin{subfigure}[b]{0.5\textwidth}
        \centering
        \begin{dot2tex}[dot,mathmode]
          graph {
            subgraph cluster_id_2 {
              label="id_2"
              style=rounded
              node [color=magenta, style=filled, shape=circle]
              i_4 [label=i_4]
              node [color=yellow, style=filled, shape=circle]
              i_1 [label=i_1]
            }
          }
        \end{dot2tex}
        \caption{The $\baseline$ clustering with cluster ids.}
    \end{subfigure}%
    ~
    \begin{subfigure}[b]{0.5\textwidth}
        \centering
        \begin{dot2tex}[dot,mathmode]
          graph {
            subgraph cluster_id_3 {
              label="id_3"
              style=rounded
              node [color=magenta, style=filled, shape=circle]
              i_4 [label=i_4]
              node [color=yellow, style=filled, shape=circle]
              i_1 [label=i_1]
            }
          }
        \end{dot2tex}
        \caption{The $\experiment$ clustering with cluster ids.}
    \end{subfigure}
    ~
    \par\bigskip
    \begin{subfigure}[t]{\textwidth}
        \centering
        \begin{dot2tex}[dot,mathmode]
          graph {
            subgraph cluster_0 {
              style=rounded
              node [color=cyan, style=filled, shape=square]
              h_3 [label=h_3]
              node [color=cyan, style=filled, shape=square]
              h_2 [label=h_2]
              node [color=yellow, style=filled, shape=square]
              h_1 [label=h_1]
            }
            subgraph cluster_1 {
              style=rounded
              node [color=magenta, style=filled, shape=circle]
              i_4 [label=i_4]
              node [color=yellow, style=filled, shape=circle]
              i_1 [label=i_1]
              node [color=magenta, style=filled, shape=square]
              h_4 [label=h_4]
            }
            subgraph cluster_2 {
              style=rounded
              node [color=red, style=filled, shape=diamond]
              id_3 [label=id_3]
            }
          }
        \end{dot2tex}
        \caption{The $\Base$ clustering.}
    \end{subfigure}
    ~
    \par\bigskip
    \begin{subfigure}[t]{\textwidth}
        \centering
        \begin{dot2tex}[dot,mathmode]
          graph {
            subgraph cluster_0 {
              style=rounded
              node [color=cyan, style=filled, shape=square]
              h_3 [label=h_3]
              node [color=cyan, style=filled, shape=square]
              h_2 [label=h_2]
              node [color=yellow, style=filled, shape=square]
              h_1 [label=h_1]
            }
            subgraph cluster_1 {
              style=rounded
              node [color=magenta, style=filled, shape=square]
              h_4 [label=h_4]
            }
            subgraph cluster_2 {
              style=rounded
              node [color=red, style=filled, shape=diamond]
              id_3 [label=id_3]
              node [color=magenta, style=filled, shape=circle]
              i_4 [label=i_4]
              node [color=yellow, style=filled, shape=circle]
              i_1 [label=i_1]
            }
          }
        \end{dot2tex}
        \caption{The $\Exp$ clustering.}
    \end{subfigure}
    ~
    \par\bigskip
    \begin{subfigure}[t]{\textwidth}
        \centering
        \begin{dot2tex}[dot,mathmode]
          graph {
            subgraph cluster_0 {
              style=rounded
              node [color=yellow, style=filled, shape=circle]
              i_1 [label=i_1]
              node [color=yellow, style=filled, shape=square]
              h_1 [label=h_1]
            }
            subgraph cluster_1 {
              style=rounded
              node [color=cyan, style=filled, shape=square]
              h_3 [label=h_3]
              node [color=cyan, style=filled, shape=square]
              h_2 [label=h_2]
            }
            subgraph cluster_2 {
              style=rounded
              node [color=magenta, style=filled, shape=circle]
              i_4 [label=i_4]
              node [color=magenta, style=filled, shape=square]
              h_4 [label=h_4]
            }
            subgraph cluster_3 {
              style=rounded
              node [color=red, style=filled, shape=diamond]
              id_3 [label=id_3]
            }
          }
        \end{dot2tex}
        \caption{The $\Ideal$ clustering.}
    \end{subfigure}
    ~
    \par\bigskip
    \makebox[\linewidth][c]{%
    \begin{subfigure}[b]{0.3666666666666667\textwidth}
        \centering
        \begin{tabular}{|l|l|}
          \hline
          $h_1$ & 1.0 \\
          \hline
          $h_2$ & 1.0 \\
          \hline
          $h_3$ & 1.0 \\
          \hline
          $h_4$ & 1.0 \\
          \hline
          $i_1$ & 1.0 \\
          \hline
          $i_4$ & 1.0 \\
          \hline
          $id_3$ & 0.001 \\
          \hline
        \end{tabular}
        \caption{The $\Weight$ mapping.}
    \end{subfigure}%
    ~
    \begin{subfigure}[b]{0.3666666666666667\textwidth}
        \centering
        \begin{tabular}{|l|r|}
          \hline
          $\JaccardDistance$ & 22.24\% \\
          \hline
          $\SplitRate$ & 22.22\% \\
          \hline
          $\MergeRate$ & 0.03\% \\
          \hline
        \end{tabular}
        \caption{Impact metrics.}
    \end{subfigure}%
    ~
    \begin{subfigure}[b]{0.3666666666666667\textwidth}
        \centering
        \begin{tabular}{|l|r|}
          \hline
          $\GoodSplitRate$ & 11.11\% \\
          \hline
          $\BadSplitRate$ & 11.11\% \\
          \hline
          $\GoodMergeRate$ & 0.00\% \\
          \hline
          $\BadMergeRate$ & 0.03\% \\
          \hline
          $\DeltaPrecision$ & 5.53\% \\
          \hline
          $\DeltaRecall$ & -16.66\% \\
          \hline
          $\IQ$ & -31.31\% \\
          \hline
        \end{tabular}
        \caption{Quality metrics.}
    \end{subfigure}
    }
    \caption{Using a fresh id for a conflated cluster.}\label{figure_fresh_id_for_conflated_cluster}
\end{figure*}

\begin{figure*}[ht!]
    \centering
    \begin{subfigure}[t]{\textwidth}
        \centering
        \begin{dot2tex}[dot,mathmode]
          graph {
            subgraph cluster_id_1 {
              label="id_1"
              style=rounded
              node [color=yellow, style=filled, shape=square]
              h_1 [label=h_1]
            }
            subgraph cluster_id_2 {
              label="id_2"
              style=rounded
              node [color=cyan, style=filled, shape=square]
              h_2 [label=h_2]
            }
            subgraph cluster_id_3 {
              label="id_3"
              style=rounded
              node [color=magenta, style=filled, shape=square]
              h_3 [label=h_3]
            }
            subgraph cluster_id_4 {
              label="id_4"
              style=rounded
              node [color=red, style=filled, shape=square]
              h_4 [label=h_4]
            }
          }
        \end{dot2tex}
        \caption{The $\hist$ clustering with cluster ids.}
    \end{subfigure}
    ~
    \par\bigskip
    \begin{subfigure}[b]{0.5\textwidth}
        \centering
        \begin{dot2tex}[dot,mathmode]
          graph {
            subgraph cluster_id_1 {
              label="id_1"
              style=rounded
              node [color=red, style=filled, shape=circle]
              i_4 [label=i_4]
              node [color=magenta, style=filled, shape=circle]
              i_3 [label=i_3]
              node [color=cyan, style=filled, shape=circle]
              i_2 [label=i_2]
              node [color=yellow, style=filled, shape=circle]
              i_1 [label=i_1]
            }
          }
        \end{dot2tex}
        \caption{The $\baseline$ clustering with cluster ids.}
    \end{subfigure}%
    ~
    \begin{subfigure}[b]{0.5\textwidth}
        \centering
        \begin{dot2tex}[dot,mathmode]
          graph {
            subgraph cluster_id_5 {
              label="id_5"
              style=rounded
              node [color=red, style=filled, shape=circle]
              i_4 [label=i_4]
              node [color=magenta, style=filled, shape=circle]
              i_3 [label=i_3]
              node [color=cyan, style=filled, shape=circle]
              i_2 [label=i_2]
              node [color=yellow, style=filled, shape=circle]
              i_1 [label=i_1]
            }
          }
        \end{dot2tex}
        \caption{The $\experiment$ clustering with cluster ids.}
    \end{subfigure}
    ~
    \par\bigskip
    \begin{subfigure}[t]{\textwidth}
        \centering
        \begin{dot2tex}[dot,mathmode]
          graph {
            subgraph cluster_0 {
              style=rounded
              node [color=red, style=filled, shape=circle]
              i_4 [label=i_4]
              node [color=magenta, style=filled, shape=circle]
              i_3 [label=i_3]
              node [color=cyan, style=filled, shape=circle]
              i_2 [label=i_2]
              node [color=yellow, style=filled, shape=circle]
              i_1 [label=i_1]
              node [color=yellow, style=filled, shape=square]
              h_1 [label=h_1]
            }
            subgraph cluster_1 {
              style=rounded
              node [color=cyan, style=filled, shape=square]
              h_2 [label=h_2]
            }
            subgraph cluster_2 {
              style=rounded
              node [color=magenta, style=filled, shape=square]
              h_3 [label=h_3]
            }
            subgraph cluster_3 {
              style=rounded
              node [color=red, style=filled, shape=square]
              h_4 [label=h_4]
            }
            subgraph cluster_4 {
              style=rounded
              node [color=green, style=filled, shape=diamond]
              id_5 [label=id_5]
            }
          }
        \end{dot2tex}
        \caption{The $\Base$ clustering.}
    \end{subfigure}
    ~
    \par\bigskip
    \begin{subfigure}[t]{\textwidth}
        \centering
        \begin{dot2tex}[dot,mathmode]
          graph {
            subgraph cluster_0 {
              style=rounded
              node [color=yellow, style=filled, shape=square]
              h_1 [label=h_1]
            }
            subgraph cluster_1 {
              style=rounded
              node [color=cyan, style=filled, shape=square]
              h_2 [label=h_2]
            }
            subgraph cluster_2 {
              style=rounded
              node [color=magenta, style=filled, shape=square]
              h_3 [label=h_3]
            }
            subgraph cluster_3 {
              style=rounded
              node [color=red, style=filled, shape=square]
              h_4 [label=h_4]
            }
            subgraph cluster_4 {
              style=rounded
              node [color=green, style=filled, shape=diamond]
              id_5 [label=id_5]
              node [color=red, style=filled, shape=circle]
              i_4 [label=i_4]
              node [color=magenta, style=filled, shape=circle]
              i_3 [label=i_3]
              node [color=cyan, style=filled, shape=circle]
              i_2 [label=i_2]
              node [color=yellow, style=filled, shape=circle]
              i_1 [label=i_1]
            }
          }
        \end{dot2tex}
        \caption{The $\Exp$ clustering.}
    \end{subfigure}
    ~
    \par\bigskip
    \begin{subfigure}[t]{\textwidth}
        \centering
        \begin{dot2tex}[dot,mathmode]
          graph {
            subgraph cluster_0 {
              style=rounded
              node [color=yellow, style=filled, shape=circle]
              i_1 [label=i_1]
              node [color=yellow, style=filled, shape=square]
              h_1 [label=h_1]
            }
            subgraph cluster_1 {
              style=rounded
              node [color=cyan, style=filled, shape=circle]
              i_2 [label=i_2]
              node [color=cyan, style=filled, shape=square]
              h_2 [label=h_2]
            }
            subgraph cluster_2 {
              style=rounded
              node [color=magenta, style=filled, shape=circle]
              i_3 [label=i_3]
              node [color=magenta, style=filled, shape=square]
              h_3 [label=h_3]
            }
            subgraph cluster_3 {
              style=rounded
              node [color=red, style=filled, shape=circle]
              i_4 [label=i_4]
              node [color=red, style=filled, shape=square]
              h_4 [label=h_4]
            }
            subgraph cluster_4 {
              style=rounded
              node [color=green, style=filled, shape=diamond]
              id_5 [label=id_5]
            }
          }
        \end{dot2tex}
        \caption{The $\Ideal$ clustering.}
    \end{subfigure}
    ~
    \par\bigskip
    \makebox[\linewidth][c]{%
    \begin{subfigure}[b]{0.3666666666666667\textwidth}
        \centering
        \begin{tabular}{|l|l|}
          \hline
          $h_1$ & 10.0 \\
          \hline
          $h_2$ & 1.0 \\
          \hline
          $h_3$ & 1.0 \\
          \hline
          $h_4$ & 1.0 \\
          \hline
          $i_1$ & 1.0 \\
          \hline
          $i_2$ & 1.0 \\
          \hline
          $i_3$ & 1.0 \\
          \hline
          $i_4$ & 1.0 \\
          \hline
          $id_5$ & 0.001 \\
          \hline
        \end{tabular}
        \caption{The $\Weight$ mapping.}
    \end{subfigure}%
    ~
    \begin{subfigure}[b]{0.3666666666666667\textwidth}
        \centering
        \begin{tabular}{|l|r|}
          \hline
          $\JaccardDistance$ & 33.62\% \\
          \hline
          $\SplitRate$ & 33.61\% \\
          \hline
          $\MergeRate$ & 0.01\% \\
          \hline
        \end{tabular}
        \caption{Impact metrics.}
    \end{subfigure}%
    ~
    \begin{subfigure}[b]{0.3666666666666667\textwidth}
        \centering
        \begin{tabular}{|l|r|}
          \hline
          $\GoodSplitRate$ & 25.21\% \\
          \hline
          $\BadSplitRate$ & 8.40\% \\
          \hline
          $\GoodMergeRate$ & 0.00\% \\
          \hline
          $\BadMergeRate$ & 0.01\% \\
          \hline
          $\DeltaPrecision$ & 12.60\% \\
          \hline
          $\DeltaRecall$ & -10.69\% \\
          \hline
          $\IQ$ & 16.07\% \\
          \hline
        \end{tabular}
        \caption{Quality metrics.}
    \end{subfigure}
    }
    \caption{Using a fresh id for a conflated cluster can be better than using an important historical id.}\label{figure_fresh_id_is_better_than_important_historical_id}
\end{figure*}

\begin{figure*}[ht!]
    \centering
    \begin{subfigure}[t]{\textwidth}
        \centering
        \begin{dot2tex}[dot,mathmode]
          graph {
            subgraph cluster_id_1 {
              label="id_1"
              style=rounded
              node [color=yellow, style=filled, shape=square]
              h_1 [label=h_1]
            }
            subgraph cluster_id_2 {
              label="id_2"
              style=rounded
              node [color=cyan, style=filled, shape=square]
              h_2 [label=h_2]
            }
          }
        \end{dot2tex}
        \caption{The $\hist$ clustering with cluster ids.}
    \end{subfigure}
    ~
    \par\bigskip
    \begin{subfigure}[b]{0.5\textwidth}
        \centering
        \begin{dot2tex}[dot,mathmode]
          graph {
            subgraph cluster_id_2 {
              label="id_2"
              style=rounded
              node [color=yellow, style=filled, shape=circle]
              i_1 [label=i_1]
            }
          }
        \end{dot2tex}
        \caption{The $\baseline$ clustering with cluster ids.}
    \end{subfigure}%
    ~
    \begin{subfigure}[b]{0.5\textwidth}
        \centering
        \begin{dot2tex}[dot,mathmode]
          graph {
            subgraph cluster_id_3 {
              label="id_3"
              style=rounded
              node [color=yellow, style=filled, shape=circle]
              i_1 [label=i_1]
            }
          }
        \end{dot2tex}
        \caption{The $\experiment$ clustering with cluster ids.}
    \end{subfigure}
    ~
    \par\bigskip
    \begin{subfigure}[t]{\textwidth}
        \centering
        \begin{dot2tex}[dot,mathmode]
          graph {
            subgraph cluster_0 {
              style=rounded
              node [color=yellow, style=filled, shape=square]
              h_1 [label=h_1]
            }
            subgraph cluster_1 {
              style=rounded
              node [color=yellow, style=filled, shape=circle]
              i_1 [label=i_1]
              node [color=cyan, style=filled, shape=square]
              h_2 [label=h_2]
            }
            subgraph cluster_2 {
              style=rounded
              node [color=magenta, style=filled, shape=diamond]
              id_3 [label=id_3]
            }
          }
        \end{dot2tex}
        \caption{The $\Base$ clustering.}
    \end{subfigure}
    ~
    \par\bigskip
    \begin{subfigure}[t]{\textwidth}
        \centering
        \begin{dot2tex}[dot,mathmode]
          graph {
            subgraph cluster_0 {
              style=rounded
              node [color=yellow, style=filled, shape=square]
              h_1 [label=h_1]
            }
            subgraph cluster_1 {
              style=rounded
              node [color=cyan, style=filled, shape=square]
              h_2 [label=h_2]
            }
            subgraph cluster_2 {
              style=rounded
              node [color=magenta, style=filled, shape=diamond]
              id_3 [label=id_3]
              node [color=yellow, style=filled, shape=circle]
              i_1 [label=i_1]
            }
          }
        \end{dot2tex}
        \caption{The $\Exp$ clustering.}
    \end{subfigure}
    ~
    \par\bigskip
    \begin{subfigure}[t]{\textwidth}
        \centering
        \begin{dot2tex}[dot,mathmode]
          graph {
            subgraph cluster_0 {
              style=rounded
              node [color=yellow, style=filled, shape=circle]
              i_1 [label=i_1]
              node [color=yellow, style=filled, shape=square]
              h_1 [label=h_1]
            }
            subgraph cluster_1 {
              style=rounded
              node [color=cyan, style=filled, shape=square]
              h_2 [label=h_2]
            }
            subgraph cluster_2 {
              style=rounded
              node [color=magenta, style=filled, shape=diamond]
              id_3 [label=id_3]
            }
          }
        \end{dot2tex}
        \caption{The $\Ideal$ clustering.}
    \end{subfigure}
    ~
    \par\bigskip
    \makebox[\linewidth][c]{%
    \begin{subfigure}[b]{0.3666666666666667\textwidth}
        \centering
        \begin{tabular}{|l|l|}
          \hline
          $h_1$ & 1.0 \\
          \hline
          $h_2$ & 1.0 \\
          \hline
          $i_1$ & 1.0 \\
          \hline
          $id_3$ & 0.001 \\
          \hline
        \end{tabular}
        \caption{The $\Weight$ mapping.}
    \end{subfigure}%
    ~
    \begin{subfigure}[b]{0.3666666666666667\textwidth}
        \centering
        \begin{tabular}{|l|r|}
          \hline
          $\JaccardDistance$ & 33.36\% \\
          \hline
          $\SplitRate$ & 33.32\% \\
          \hline
          $\MergeRate$ & 0.07\% \\
          \hline
        \end{tabular}
        \caption{Impact metrics.}
    \end{subfigure}%
    ~
    \begin{subfigure}[b]{0.3666666666666667\textwidth}
        \centering
        \begin{tabular}{|l|r|}
          \hline
          $\GoodSplitRate$ & 33.32\% \\
          \hline
          $\BadSplitRate$ & 0.00\% \\
          \hline
          $\GoodMergeRate$ & 0.00\% \\
          \hline
          $\BadMergeRate$ & 0.07\% \\
          \hline
          $\DeltaPrecision$ & 33.26\% \\
          \hline
          $\DeltaRecall$ & 0.00\% \\
          \hline
          $\IQ$ & 66.46\% \\
          \hline
        \end{tabular}
        \caption{Quality metrics.}
    \end{subfigure}
    }
    \caption{Instead of a totally wrong id, it's better to use a fresh id.}\label{figure_better_to_use_fresh_id_than_wrong_id}
\end{figure*}

\begin{figure*}[ht!]
    \centering
    \begin{subfigure}[t]{\textwidth}
        \centering
        \begin{dot2tex}[dot,mathmode]
          graph {
            subgraph cluster_id_1 {
              label="id_1"
              style=rounded
              node [color=yellow, style=filled, shape=square]
              h_1 [label=h_1]
            }
            subgraph cluster_id_2 {
              label="id_2"
              style=rounded
              node [color=cyan, style=filled, shape=square]
              h_2 [label=h_2]
            }
          }
        \end{dot2tex}
        \caption{The $\hist$ clustering with cluster ids.}
    \end{subfigure}
    ~
    \par\bigskip
    \begin{subfigure}[b]{0.5\textwidth}
        \centering
        \begin{dot2tex}[dot,mathmode]
          graph {
            subgraph cluster_id_3 {
              label="id_3"
              style=rounded
              node [color=yellow, style=filled, shape=circle]
              i_1 [label=i_1]
            }
            subgraph cluster_id_4 {
              label="id_4"
              style=rounded
              node [color=cyan, style=filled, shape=circle]
              i_2 [label=i_2]
            }
          }
        \end{dot2tex}
        \caption{The $\baseline$ clustering with cluster ids.}
    \end{subfigure}%
    ~
    \begin{subfigure}[b]{0.5\textwidth}
        \centering
        \begin{dot2tex}[dot,mathmode]
          graph {
            subgraph cluster_id_5 {
              label="id_5"
              style=rounded
              node [color=yellow, style=filled, shape=circle]
              i_1 [label=i_1]
            }
            subgraph cluster_id_6 {
              label="id_6"
              style=rounded
              node [color=cyan, style=filled, shape=circle]
              i_2 [label=i_2]
            }
          }
        \end{dot2tex}
        \caption{The $\experiment$ clustering with cluster ids.}
    \end{subfigure}
    ~
    \par\bigskip
    \begin{subfigure}[t]{\textwidth}
        \centering
        \begin{dot2tex}[dot,mathmode]
          graph {
            subgraph cluster_0 {
              style=rounded
              node [color=yellow, style=filled, shape=square]
              h_1 [label=h_1]
            }
            subgraph cluster_1 {
              style=rounded
              node [color=cyan, style=filled, shape=square]
              h_2 [label=h_2]
            }
            subgraph cluster_2 {
              style=rounded
              node [color=magenta, style=filled, shape=diamond]
              id_3 [label=id_3]
              node [color=yellow, style=filled, shape=circle]
              i_1 [label=i_1]
            }
            subgraph cluster_3 {
              style=rounded
              node [color=red, style=filled, shape=diamond]
              id_4 [label=id_4]
              node [color=cyan, style=filled, shape=circle]
              i_2 [label=i_2]
            }
            subgraph cluster_4 {
              style=rounded
              node [color=green, style=filled, shape=diamond]
              id_5 [label=id_5]
            }
            subgraph cluster_5 {
              style=rounded
              node [color=orange, style=filled, shape=diamond]
              id_6 [label=id_6]
            }
          }
        \end{dot2tex}
        \caption{The $\Base$ clustering.}
    \end{subfigure}
    ~
    \par\bigskip
    \begin{subfigure}[t]{\textwidth}
        \centering
        \begin{dot2tex}[dot,mathmode]
          graph {
            subgraph cluster_0 {
              style=rounded
              node [color=yellow, style=filled, shape=square]
              h_1 [label=h_1]
            }
            subgraph cluster_1 {
              style=rounded
              node [color=cyan, style=filled, shape=square]
              h_2 [label=h_2]
            }
            subgraph cluster_2 {
              style=rounded
              node [color=green, style=filled, shape=diamond]
              id_5 [label=id_5]
              node [color=yellow, style=filled, shape=circle]
              i_1 [label=i_1]
            }
            subgraph cluster_3 {
              style=rounded
              node [color=orange, style=filled, shape=diamond]
              id_6 [label=id_6]
              node [color=cyan, style=filled, shape=circle]
              i_2 [label=i_2]
            }
            subgraph cluster_4 {
              style=rounded
              node [color=magenta, style=filled, shape=diamond]
              id_3 [label=id_3]
            }
            subgraph cluster_5 {
              style=rounded
              node [color=red, style=filled, shape=diamond]
              id_4 [label=id_4]
            }
          }
        \end{dot2tex}
        \caption{The $\Exp$ clustering.}
    \end{subfigure}
    ~
    \par\bigskip
    \begin{subfigure}[t]{\textwidth}
        \centering
        \begin{dot2tex}[dot,mathmode]
          graph {
            subgraph cluster_0 {
              style=rounded
              node [color=yellow, style=filled, shape=circle]
              i_1 [label=i_1]
              node [color=yellow, style=filled, shape=square]
              h_1 [label=h_1]
            }
            subgraph cluster_1 {
              style=rounded
              node [color=cyan, style=filled, shape=circle]
              i_2 [label=i_2]
              node [color=cyan, style=filled, shape=square]
              h_2 [label=h_2]
            }
            subgraph cluster_2 {
              style=rounded
              node [color=magenta, style=filled, shape=diamond]
              id_3 [label=id_3]
            }
            subgraph cluster_3 {
              style=rounded
              node [color=red, style=filled, shape=diamond]
              id_4 [label=id_4]
            }
            subgraph cluster_4 {
              style=rounded
              node [color=green, style=filled, shape=diamond]
              id_5 [label=id_5]
            }
            subgraph cluster_5 {
              style=rounded
              node [color=orange, style=filled, shape=diamond]
              id_6 [label=id_6]
            }
          }
        \end{dot2tex}
        \caption{The $\Ideal$ clustering.}
    \end{subfigure}
    ~
    \par\bigskip
    \makebox[\linewidth][c]{%
    \begin{subfigure}[b]{0.3666666666666667\textwidth}
        \centering
        \begin{tabular}{|l|l|}
          \hline
          $h_1$ & 1.0 \\
          \hline
          $h_2$ & 1.0 \\
          \hline
          $i_1$ & 1.0 \\
          \hline
          $i_2$ & 1.0 \\
          \hline
          $id_3$ & 0.001 \\
          \hline
          $id_4$ & 0.001 \\
          \hline
          $id_5$ & 0.001 \\
          \hline
          $id_6$ & 0.001 \\
          \hline
        \end{tabular}
        \caption{The $\Weight$ mapping.}
    \end{subfigure}%
    ~
    \begin{subfigure}[b]{0.3666666666666667\textwidth}
        \centering
        \begin{tabular}{|l|r|}
          \hline
          $\JaccardDistance$ & 0.20\% \\
          \hline
          $\SplitRate$ & 0.10\% \\
          \hline
          $\MergeRate$ & 0.10\% \\
          \hline
        \end{tabular}
        \caption{Impact metrics.}
    \end{subfigure}%
    ~
    \begin{subfigure}[b]{0.3666666666666667\textwidth}
        \centering
        \begin{tabular}{|l|r|}
          \hline
          $\GoodSplitRate$ & 0.10\% \\
          \hline
          $\BadSplitRate$ & 0.00\% \\
          \hline
          $\GoodMergeRate$ & 0.00\% \\
          \hline
          $\BadMergeRate$ & 0.10\% \\
          \hline
          $\DeltaPrecision$ & 0.00\% \\
          \hline
          $\DeltaRecall$ & 0.00\% \\
          \hline
          $\IQ$ & 0.00\% \\
          \hline
        \end{tabular}
        \caption{Quality metrics.}
    \end{subfigure}
    }
    \caption{The baseline and experiment assign fresh ids to all clusters.}\label{figure_base_and_exp_use_only_fresh_ids}
\end{figure*}

\subsection{Discussion}

The ABCDE impact metrics are pointwise metrics~\cite{vanstaden2024pointwise} for which it makes no difference whether a cluster member is an ``atom'' or actually many smaller members that move together and whose total weight is equal to that of the ``atom''. So the metrics that measure the diff Impact will be the same as for the setting with slightly modified definitions:
\begin{itemize}
\item $\HistMembersOrId(id) = \{id\}$
\item $\Weight$ associates each $id \in \AllIds$ with $\max(\weight(\Cluster_\hist(id)), k)$. 
\end{itemize}
The reason why we expand the historical items associated with a historical cluster id is because that helps with the quality metrics. In particular:
\begin{itemize}
\item It simplifies the questions for human judgement. \\ Asking whether two items share the same identity is easy, while asking whether a current item would be a suitable a member of a historical cluster can be hard/impossible to answer when the historical cluster is conflated and contains items with various identities.
\item It leads to more nuanced metrics. \\ Instead of a binary verdict of whether a current item would be a suitable member of a historical cluster, we can obtain the degree to which the item shares the identity with the historical cluster by considering its members individually.
\end{itemize}

\section{Generalizations}

The basic evaluation setup of the previous section can be generalized in several ways.

\subsection{Changing cluster memberships and cluster ids simultaneously}

The basic setup evaluates the relative merits of two schemes that assign ids to a fixed clustering $C$.
Conceptually there is a clustering step, followed by an id assignment step which can be done in two ways, and the id assignment step is evaluated (the clustering step can be evaluated with vanilla ABCDE).

The definitions from before generalize nicely to the setting where id assignment is not a separate step. This has interesting applications, because it allows us to evaluate operations that change some cluster memberships and some cluster ids in one go, and where it is unnatural/unnecessary to decompose it into two steps for evaluation purposes\footnote{A decomposition might yield additional insight, and can still be performed on demand, but it is cumbersome to insist on a decomposition every time.}.

In this generalization, we have:
\begin{itemize}
\item
A current clustering $C_\baseline$, with ids assigned by the baseline. \\
A current clustering $C_\experiment$, with ids assigned by the experiment. \\
Clusterings $C_\baseline$ and $C_\experiment$ involve the same items and item data.
\item
$\Ids_\baseline$ denotes the set of ids assigned by the baseline to clustering $C_\baseline$. \\
$\Ids_\experiment$ denotes the set of ids assigned by the experiment to clustering $C_\experiment$.
\item 
The weight mapping for the current items will depend the weight assignment scheme being used. Weight assignment schemes are discussed in Section 3.1 of~\cite{vanstadengrubb2024abcde}. If the scheme uses the intrinsic weight of an item, then each current item already has an unambiguous weight. If the scheme uses weights that depend on the clustering, then each current item will have two weights (one in the baseline and one in the experiment), which can be combined with $\max$ to obtain a single weight value.
\end{itemize}
All the other definitions stay the same as before.

Figure~\ref{figure_fixing_cluster_memberships_and_ids_in_one_go} shows an example where the experiment fixes the cluster memberships and id assignments in one go.

\begin{figure*}[ht!]
    \centering
    \begin{subfigure}[t]{\textwidth}
        \centering
        \begin{dot2tex}[dot,mathmode]
          graph {
            subgraph cluster_id_1 {
              label="id_1"
              style=rounded
              node [color=yellow, style=filled, shape=square]
              h_2 [label=h_2]
              node [color=yellow, style=filled, shape=square]
              h_1 [label=h_1]
            }
            subgraph cluster_id_2 {
              label="id_2"
              style=rounded
              node [color=cyan, style=filled, shape=square]
              h_3 [label=h_3]
            }
          }
        \end{dot2tex}
        \caption{The $\hist$ clustering with cluster ids.}
    \end{subfigure}
    ~
    \par\bigskip
    \begin{subfigure}[b]{0.5\textwidth}
        \centering
        \begin{dot2tex}[dot,mathmode]
          graph {
            subgraph cluster_id_2 {
              label="id_2"
              style=rounded
              node [color=cyan, style=filled, shape=circle]
              i_3 [label=i_3]
              node [color=yellow, style=filled, shape=circle]
              i_1 [label=i_1]
            }
            subgraph cluster_id_1 {
              label="id_1"
              style=rounded
              node [color=yellow, style=filled, shape=circle]
              i_2 [label=i_2]
            }
          }
        \end{dot2tex}
        \caption{The $\baseline$ clustering with cluster ids.}
    \end{subfigure}%
    ~
    \begin{subfigure}[b]{0.5\textwidth}
        \centering
        \begin{dot2tex}[dot,mathmode]
          graph {
            subgraph cluster_id_1 {
              label="id_1"
              style=rounded
              node [color=yellow, style=filled, shape=circle]
              i_2 [label=i_2]
              node [color=yellow, style=filled, shape=circle]
              i_1 [label=i_1]
            }
            subgraph cluster_id_2 {
              label="id_2"
              style=rounded
              node [color=cyan, style=filled, shape=circle]
              i_3 [label=i_3]
            }
          }
        \end{dot2tex}
        \caption{The $\experiment$ clustering with cluster ids.}
    \end{subfigure}
    ~
    \par\bigskip
    \begin{subfigure}[t]{\textwidth}
        \centering
        \begin{dot2tex}[dot,mathmode]
          graph {
            subgraph cluster_0 {
              style=rounded
              node [color=yellow, style=filled, shape=circle]
              i_2 [label=i_2]
              node [color=yellow, style=filled, shape=square]
              h_2 [label=h_2]
              node [color=yellow, style=filled, shape=square]
              h_1 [label=h_1]
            }
            subgraph cluster_1 {
              style=rounded
              node [color=cyan, style=filled, shape=circle]
              i_3 [label=i_3]
              node [color=yellow, style=filled, shape=circle]
              i_1 [label=i_1]
              node [color=cyan, style=filled, shape=square]
              h_3 [label=h_3]
            }
          }
        \end{dot2tex}
        \caption{The $\Base$ clustering.}
    \end{subfigure}
    ~
    \par\bigskip
    \begin{subfigure}[t]{\textwidth}
        \centering
        \begin{dot2tex}[dot,mathmode]
          graph {
            subgraph cluster_0 {
              style=rounded
              node [color=yellow, style=filled, shape=circle]
              i_2 [label=i_2]
              node [color=yellow, style=filled, shape=circle]
              i_1 [label=i_1]
              node [color=yellow, style=filled, shape=square]
              h_2 [label=h_2]
              node [color=yellow, style=filled, shape=square]
              h_1 [label=h_1]
            }
            subgraph cluster_1 {
              style=rounded
              node [color=cyan, style=filled, shape=circle]
              i_3 [label=i_3]
              node [color=cyan, style=filled, shape=square]
              h_3 [label=h_3]
            }
          }
        \end{dot2tex}
        \caption{The $\Exp$ clustering.}
    \end{subfigure}
    ~
    \par\bigskip
    \begin{subfigure}[t]{\textwidth}
        \centering
        \begin{dot2tex}[dot,mathmode]
          graph {
            subgraph cluster_0 {
              style=rounded
              node [color=yellow, style=filled, shape=circle]
              i_2 [label=i_2]
              node [color=yellow, style=filled, shape=circle]
              i_1 [label=i_1]
              node [color=yellow, style=filled, shape=square]
              h_2 [label=h_2]
              node [color=yellow, style=filled, shape=square]
              h_1 [label=h_1]
            }
            subgraph cluster_1 {
              style=rounded
              node [color=cyan, style=filled, shape=circle]
              i_3 [label=i_3]
              node [color=cyan, style=filled, shape=square]
              h_3 [label=h_3]
            }
          }
        \end{dot2tex}
        \caption{The $\Ideal$ clustering.}
    \end{subfigure}
    ~
    \par\bigskip
    \makebox[\linewidth][c]{%
    \begin{subfigure}[b]{0.3666666666666667\textwidth}
        \centering
        \begin{tabular}{|l|l|}
          \hline
          $h_1$ & 1.0 \\
          \hline
          $h_2$ & 1.0 \\
          \hline
          $h_3$ & 1.0 \\
          \hline
          $i_1$ & 1.0 \\
          \hline
          $i_2$ & 1.0 \\
          \hline
          $i_3$ & 1.0 \\
          \hline
        \end{tabular}
        \caption{The $\Weight$ mapping.}
    \end{subfigure}%
    ~
    \begin{subfigure}[b]{0.3666666666666667\textwidth}
        \centering
        \begin{tabular}{|l|r|}
          \hline
          $\JaccardDistance$ & 37.50\% \\
          \hline
          $\SplitRate$ & 22.22\% \\
          \hline
          $\MergeRate$ & 25.00\% \\
          \hline
        \end{tabular}
        \caption{Impact metrics.}
    \end{subfigure}%
    ~
    \begin{subfigure}[b]{0.3666666666666667\textwidth}
        \centering
        \begin{tabular}{|l|r|}
          \hline
          $\GoodSplitRate$ & 22.22\% \\
          \hline
          $\BadSplitRate$ & 0.00\% \\
          \hline
          $\GoodMergeRate$ & 25.00\% \\
          \hline
          $\BadMergeRate$ & 0.00\% \\
          \hline
          $\DeltaPrecision$ & 22.22\% \\
          \hline
          $\DeltaRecall$ & 25.00\% \\
          \hline
          $\IQ$ & 100.00\% \\
          \hline
        \end{tabular}
        \caption{Quality metrics.}
    \end{subfigure}
    }
    \caption{Fixing cluster memberships and ids in one go.}\label{figure_fixing_cluster_memberships_and_ids_in_one_go}
\end{figure*}

\subsection{Current clusterings over different items}

In practice it can be useful for some applications to evaluate an experiment with respect to a baseline that did not use identical input data. For example, if the item data changes daily, then the baseline and experiment clusterings can be two successive daily outputs of the clustering system, and they can be evaluated with respect to a historical clustering of say one week or one month ago. Such a setup can supply useful information about the day-to-day magnitude of the differences in cluster membership and cluster id assignment.

For such cases we can assume that both the baseline and experiment use current items. To obtain a single set of items that was clustered by the baseline and experiment, we use the standard mechanism of ABCDE (and of evaluation with respect to a ground truth clustering): intersect the items of the baseline and experiment, restrict clusters to the resulting items, and remove empty clusters (and their cluster ids) from consideration. The weight of a remaining item is taken to be the $\max$ of its weight in the baseline and experiment. We can use the generalization from the previous section to account for the fact that the baseline and experiment clusters are different. For the human judgements, we can decide whether to use the baseline's item data or the experiment's item data when a current item is involved; to keep things simple we can always use the experiment's item data and assume that it didn't change much since the baseline.

Standard evaluations to compare different id assignment schemes should not use this setup, because the differences in the items and their data can and do have an effect on both cluster membership and id assignment. Comparisons between schemes are best performed sans such effects.

\subsection{Putting emphasis on the present or the past}

The primary knob to influence ABCDE metrics is the weights of the items. It is desirable to use an application's usual weight assignment scheme to derive weights for the historical items and the current items. However, after doing that we are still free to scale the weights of the historical items by a constant positive factor, denoted by $\mathit{HistScaleFactor}$. The basic setup uses $\mathit{HistScaleFactor} = 1$. Using $\mathit{HistScaleFactor} < 1$ will put more emphasis on the present, i.e. the current situation/items, while using $\mathit{HistScaleFactor} > 1$ will put more emphasis on the past, i.e. the historical situation/items.

Aside: we can alternatively scale the weights of the current items. But then we have to remember to scale the weight of each $id \in \NonHistIds$ as well, because conceptually these also belong to the `current' situation. The formulation with $\mathit{HistScaleFactor}$ is simpler.

\subsection{Considering multiple historical clusterings}

We can generalize the definitions to take multiple historical clusterings into account. Using that in practice will increase the computational overhead, and more human ratings will be needed to get reasonable confidence intervals. But it could make the measurements more robust, because there are more historical cluster ids and each historical id has more context.

Suppose we have $n$ historical epochs, where each epoch is a clustering of a set of items, together with weights for the items and ids for the clusters. Notation: for epoch $j \in 1 \ldots n$,
\begin{itemize}
\item $H_j$ denotes the historical clustering of epoch $j$ of the items in $\mathit{Items}_{H_j}$. These historical items are considered to be disjoint from the items of all other epochs.
\item $\weight_{H_j}(h)$ associates each historical item $h \in \mathit{Items}_{H_j}$ with a positive real weight.
\item $\Ids_{H_j}$ is the set of all ids of the clusters of $H_j$.
\item $\Cluster_{H_j}(id)$ denotes the cluster (i.e. set of items) that $H_j$ associated with $id$. If $id \notin \Ids_{H_j}$, then $\Cluster_{H_j}(id) = \emptyset$.
\end{itemize}
Moreover, suppose that each historical epoch $j$ has a positive real scalar $\mathit{Weight}H_j$, which indicates its relative importance with respect to the other historical epochs.

From this we can construct a single historical clustering, whose items have weights and whose clusters have ids, as follows:
\begin{itemize}
\item $\mathit{Items}_\hist = \bigcup\{\mathit{Items}_{H_j} | j \in 1 \ldots n\}$ \\
So the historical clustering involves all items from all $n$ historical epochs, and items from different epochs are kept separate.
\item $\Ids_\hist = \bigcup\{\Ids_{H_j} | j \in 1 \ldots n\}$
\item For each $id \in \Ids_\hist$: \\
$\Cluster_\hist(id) = \bigcup\{\Cluster_{H_j}(id) | j \in 1 \ldots n\}$
\item It remains to define the weights of the historical items. \\
For each $h \in \mathit{Items}_\hist$ there exists a unique $H_j$ such that $h \in \mathit{Items}_{H_j}$, and we define: \\
$$\weight(h) = \mathit{ScaleFactor}{H_j} \cdot \weight_{H_j}(h)$$
where
$$\mathit{ScaleFactor}{H_j} = \frac{\mathit{Weight}H_j}{\sum_{l = 1}^n \mathit{Weight}H_l}$$
\end{itemize}

Note that this generalization treats an item from a historical epoch as separate from the items in other historical epochs and also from current items. But that has no bearing on the computation of the metrics, which can proceed in the same way as before. Note also that it is a proper generalization, because the special case where there is only one historical epoch will yield the same $\hist$ clustering as in the basic setup, and hence also the same metrics.

\section{Conclusion}

This paper considers the problem of evaluating the relative merits of cluster id assignment schemes. To do that, it transforms the problem of cluster id assignment into a problem of cluster membership, which is in turn evaluated with ABCDE. As a result, we get Impact metrics that characterize the magnitude of the id assignment diffs, and Quality metrics that characterize the quality of the id assignment diffs. We also get other functionality for free, for example the ability to explore the impact interactively as described in Section~4.1.1 of~\cite{vanstadengrubb2024abcde}, the scalability of evaluation to billions of items and millions of clusters, and the ability to use different importance values for different items. The paper also describes several generalizations to the basic evaluation setup for id assignment schemes. For example, it is fairly straightforward to evaluate changes that simultaneously mutate cluster memberships and cluster ids.

\bibliographystyle{plain}
\bibliography{main}

\end{document}